\documentclass[12pt,a4paper]{article}

\usepackage{bbm}
\usepackage{stmaryrd}
\usepackage{amsmath}
\usepackage{amssymb}
\usepackage{hyperref}
\usepackage[mathcal]{euscript}

\newcounter{thMM}[section]
\newcounter{leMM}[section]
\newcounter{deFF}[section]
\newcounter{exMP}[section]
\newcounter{prOP}[section]
\newcounter{coRR}[section]
\newenvironment{theorem}[1][Theorem]{\refstepcounter{thMM}\trivlist
   \item[{\bf #1~\thesection.\arabic{thMM}.}]\it\hskip3pt}{\endtrivlist}
\newenvironment{lemma}[1][Lemma]{\refstepcounter{leMM}\trivlist
   \item[{\bf #1~\thesection.\arabic{leMM}.}]\it\hskip3pt}{\endtrivlist}
\newenvironment{definition}[1][Definition]{\refstepcounter{deFF}\trivlist
   \item[{\bf #1~\thesection.\arabic{deFF}.}]\rm\hskip3pt}{\endtrivlist}
\newenvironment{example}[1][Example]{\refstepcounter{exMP}\trivlist
   \item[{\bf #1~\thesection.\arabic{exMP}.}]\rm\hskip3pt}{\endtrivlist}
\newenvironment{proposition}[1][Proposition]{\refstepcounter{prOP}\trivlist
   \item[{\bf #1~\thesection.\arabic{prOP}.}]\it\hskip3pt}{\endtrivlist}
\newenvironment{corollary}[1][Corollary]{\refstepcounter{coRR}\trivlist
   \item[{\bf #1~\thesection.\arabic{coRR}.}]\it\hskip3pt}{\endtrivlist}

\newenvironment{proof}[1][Proof]{\begin{trivlist}
\item[\hskip \labelsep {\bfseries #1}]}{\end{trivlist}}
\newenvironment{remark}[1][Remark]{\begin{trivlist}
\item[\hskip \labelsep {\bfseries #1}]}{\end{trivlist}}

\newcommand{\SN}[1]{\llbracket   #1 \rrbracket}
\newcommand{\proofend}{\flushright $\square$}
\newcommand{\Q}{\mathcal{Q}}
\newcommand{\Id}{\mathbbmss{1}}

\DeclareMathOperator{\Vect}{Vect}

\makeatletter
\renewcommand\@biblabel[1]{#1.}
\makeatother

\begin{document}
\author{Andrew James Bruce\\ 
\small{ \emph{email:} \texttt{andrewjamesbruce@googlemail.com}}}
\date{\today}
\title{Odd Jacobi structures and classical BV-gauge systems}
\maketitle

\begin{abstract}
In this paper we define   Grassmann odd analogues of  Jacobi structures on  supermanifolds. We then examine their potential use in the Batalin--Vilkovisky formalism of classical gauge theories.
\end{abstract}

\begin{small}
\noindent \textbf{Mathematics Subject Classification (2010)}. 17B70, 53D17, 58A50, 83C47.\\
\noindent \textbf{Keywords}. supermanifolds, Jacobi structures, BV-formalism.
\end{small}
\section{Introduction}
Lichnerowicz \cite{Lichnerowicz1977} introduced the notion of a Poisson manifold as well as a Jacobi manifold. Such  manifolds have found applications in classical mechanics and play an important role in quantisation. In this paper we introduce the notion of  (Grassmann) odd Jacobi (super)manifolds. The main motivation for doing so is the fact that odd Poisson brackets, also known as Schouten brackets or in the physics literature as antibrackets have found  important application in the Batalin--Vilkovisky antifield formalism of gauge theories \cite{Batalin:1981jr,Batalin:1984jr}.\\

Let us briefly recall the definition of a (classical or even) Jacobi manifold\footnote{This can all be generalised to supermanifolds, however doing so is inessential for this paper.} as a manifold equipped with a bi-vector $\Gamma$ and a vector field $E$ such that

\begin{equation}\label{even jacobi}
\SN{\Gamma, \Gamma}_{SN} =  2 E \wedge \Gamma  \hspace{30pt} L_{E}\Gamma = \SN{E,\Gamma}_{SN} =0,
\end{equation}

where $\SN{\bullet, \bullet}_{SN}$ is the Schouten--Nijenhuis bracket. (There is some leeway here in  signs due to  conventions). Given these structures one builds a Lie algebra on $C^{\infty}(M)$ viz

\begin{equation}
\{ f,g \}_{J} = \Gamma(df, dg) + fE(g) - E(f)g.
\end{equation}

Importantly this  Jacobi bracket is not a derivation unless $E =0$, which reduces the theory to that of Poisson geometry. That is if $E \neq 0$ the Leibniz rule is not exactly satisfied but contains a correction proportional to $\{f, \Id  \}_{J}$.   For a modern review of Jacobi structures, including $\mathbb{Z}$-graded versions and their relation to Lie algebroids see \cite{Grabowski2001,Grabowski2003}.\\

In  this paper we  construct a Grassmann odd analogue of Jacobi structures on supermanifolds\footnote{We refrain from calling odd Jacobi structures \emph{Schouten--Jacobi structures} as this name has been already allocated in \cite{Grabowski2001} to mean something specific and not identical to constructions found in this work.}.  \\

Studying  geometric structures on supermanifolds provides a method of confirming the philosophy that \emph{supermanifolds can  informally be thought of as ``manifolds" with commuting and anticommuting coordinates}. More than this, the inclusion of Grassmann odd degrees of freedom allow one to construct geometric structures that have no classical analogue on manifolds. These ``odd structures" such are of mathematical interest and often find applications in physics. For example, odd symplectic and Schouten structures are at the heart of the  Batalin--Vilkovisky antifield formalism.\\

The theory of Schouten manifolds and Q-manifolds (supermanifolds with a homological vector field) are both well established within the mathematical physics literature. However, construction of an odd analogue of classical Jacobi structures appears to have been overlooked. The expected properties of such a construction cannot be taken for granted.  In this work we preform  an  examination of  the basic properties of odd Jacobi structures.  \\

 The theory of odd Jacobi structures is described by ``almost Schouten structure"\newline $S \in C^{\infty}(T^{*}M)$, that is a Grassmann odd  fibre-wise polynomial of degree two (``odd Hamiltonian") and a homological vector field $Q \in \Vect(M)$, together with natural conditions analogous to Eqn.(\ref{even jacobi}). The associated odd Jacobi brackets   satisfy the standard properties of  Schouten  brackets, with the exception of the derivation property. The majority of this paper is devoted to making the previous statement concrete and exploring the elementary properties of supermanifolds equipped with odd Jacobi structures.  \\

It is then natural to wonder if odd Jacobi brackets (to be defined) can find application in the Batalin--Vilkovisky formalism.  However, we will ignore the extra gradings of ghost, antifield, antighost  number etc. as required in a detailed exploration and explore only the classical aspects. For the purposes of this paper the lack of extra gradings will not cause problems as our description of the Batalin--Vilkovisky formalism will be very elementary. We do this to simplify our work, as well as making this paper as accessible as possible.  \\

To that aim, let us sketch the main mathematical features of the classical Batalin--Vilkovisky formalism. The extended configuration space of a classical gauge theory is understood to be a supermanifold  $M$ (space of fields, ghost, antifields, antighosts etc.) equipped with a Schouten bracket, together with a Maurer--Cartan element known as the ``extended classical action" which is an even function $s \in C^{\infty}(M)$. If we denote the Schouten bracket as $\SN{\bullet, \bullet}$ then the ``extended classical action" is a solution to the ``classical master equation" (together with boundary conditions which need not concern us)

\begin{equation}\nonumber
\SN{s,s} =0.
\end{equation}

This Maurer--Cartan element governs the infinitesimal BRST-gauge symmetry of the theory. Specifically the BRST-gauge transformations are infinitesimal canonical transformations generated by $s$. That is, we have a vector field known as the classical BRST operator given by $\delta_{s} = \SN{s, \bullet}$ which encodes the gauge structure of the theory. Importantly, the classical BRST operator is nilpotent which is guaranteed by the fact that the ``extended classical action" is a Maurer-Cartan element. \\

 In this paper we show that the classical BV-antifield formalism appears to carry over to odd Jacobi manifolds with the  proviso that the ``extended classical action" not only be a Maurer--Cartan  element (with respect to the odd Jacobi brackets) but in addition be Q-closed. Such a condition need not be stipulated on Schouten manifolds as one can think of these as odd Jacobi manifolds with the  homological vector field being identically zero.   \\

\noindent \textbf{Preliminaries} \\
All vector spaces and algebras will be $\mathbb{Z}_{2}$-graded.   We will generally  omit the prefix \emph{super}. By \emph{manifold} we will mean a \emph{smooth supermanifold}. We denote the Grassmann parity of an object by \emph{tilde}: $\widetilde{A} \in \mathbb{Z}_{2}$. By \emph{even} or \emph{odd} we will be referring explicitly to the Grassmann parity.\\

 A \emph{Poisson} $(\varepsilon = 0)$  or \emph{Schouten} $(\varepsilon = 1)$ \emph{algebra} is understood as a vector space $\mathbb{A}$ with a bilinear associative multiplication and a bilinear operation $\{\bullet , \bullet\}_{\varepsilon}: \mathbb{A}  \otimes \mathbb{A} \rightarrow \mathbb{A}$ such that:
\begin{list}{}
\item \textbf{Grading} $\widetilde{\{a,b \}_{\varepsilon}} = \widetilde{a} + \widetilde{b} + \varepsilon$
\item \textbf{Skewsymmetry} $\{a,b\}_{\varepsilon} = -(-1)^{(\tilde{a}+ \varepsilon)(\tilde{b}+ \varepsilon)} \{b,a \}_{\varepsilon}$
\item \textbf{Jacobi Identity} $\displaystyle\sum\limits_{\textnormal{cyclic}(a,b,c)} (-1)^{(\tilde{a}+ \varepsilon)(\tilde{c}+ \varepsilon)}\{a,\{b,c\}_{\varepsilon}  \}_{\varepsilon}= 0$
\item \textbf{Leibniz Rule} $\{a,bc \}_{\varepsilon} = \{a,b \}_{\varepsilon}c + (-1)^{(\tilde{a} + \varepsilon)\tilde{b}} b \{a,c \}_{\varepsilon}$
\end{list} \vspace{10pt}
for all homogenous elements $a,b,c \in \mathbb{A}$.\\

If the Leibniz rule does not hold identically, but is modified as
\begin{equation}\nonumber
\{a,bc \}_{\varepsilon} = \{a,b \}_{\varepsilon}c + (-1)^{(\tilde{a} + \varepsilon)\tilde{b}} b \{a,c \}_{\varepsilon} - \{a ,\Id  \} bc,
\end{equation}

then we have \emph{even} ($\epsilon = 0)$ or \emph{odd} ($\epsilon = 1)$ \emph{Jacobi algebras}.\\

A manifold $M$ such that $C^{\infty}(M)$ is a Poisson/Schouten algebra is known as a \emph{Poisson/Schouten manifold}. In particular the cotangent of a manifold comes equipped with a canonical Poisson structure.\\

Let us employ   natural local coordinates $(x^{A}, p_{A})$ on $T^{*}M$, with $\widetilde{x}^{A} = \widetilde{A}$ and $\widetilde{p}_{A} = \widetilde{A}$. Local diffeomorphisms on $M$ induce vector  bundle automorphism on $T^{*}M$ of the form
\begin{equation}\nonumber
\overline{x}^{A} = \overline{x}^{A}(x), \hspace{30pt} \overline{p}_{A}  = \left(\frac{\partial x^{B}}{\partial \overline{x}^{A}}\right)p_{B}.
\end{equation}

We will in effect use the local description as a \emph{natural vector bundle} to define the cotangent bundle of a supermanifold.  The canonical Poisson bracket on the cotangent is given by

\begin{equation}\nonumber
\{ F,G \} = (-1)^{\widetilde{A} \widetilde{F} + \widetilde{A}} \frac{\partial F}{\partial p_{A}}\frac{\partial G}{\partial x^{A}} - (-1)^{\widetilde{A}\widetilde{F}}\frac{\partial  F}{\partial x^{A}} \frac{\partial G}{\partial p_{A}}.
\end{equation}\\

A manifold equipped with an odd vector field $Q$, such that the non-trivial condition $Q^{2}= \frac{1}{2}[Q,Q]=0$ holds, is known as a \emph{Q-manifold} and the vector field $Q$ is known as a \emph{homological vector field} for obvious reasons. \\
\newpage 
\noindent \textbf{Organisation of paper}\\
In \S(\ref{odd jacobi}) we define odd Jacobi structures, odd Jacobi manifolds and explore their elementary properties. In \S(\ref{MC elements}) we begin to explore potential application in gauge theory. We end this paper with some concluding remarks in \S(\ref{concluding remarks}).\\

\section{Odd Jacobi structures}\label{odd jacobi}
In this section we define odd Jacobi manifolds and show that much of the theory of classical Jacobi manifolds carries over to the odd case. One should of course keep in mind the similarities and difference with Schouten manifolds. \\
\begin{definition}\label{def odd jacobi}
An \textbf{odd Jacobi structure} $ J := (S,Q)$ on a manifold $M$  consists of
\begin{itemize}
\item an odd function $S \in C^{\infty}(T^{*}M)$, of degree two in fibre coordinates,
\item an odd vector field $Q \in \Vect(M)$,
\end{itemize}
such that the following conditions hold:
\begin{enumerate}
\item the homological condition $Q^{2} = \frac{1}{2} [Q,Q]=0$,
\item the invariance condition  $L_{Q}S = 0$,
\item the compatibility condition $\{S,S \}= - 2 \Q S $,
\end{enumerate}
 Here $\Q \in C^{\infty}(T^{*}M)$ is the principle symbol or ``Hamiltonian"  of the vector field $Q$. The brackets $\{ \bullet, \bullet \}$ are the canonical Poisson brackets on the cotangent bundle of the manifold.
\end{definition}

\begin{remark}
Note  $[E,E]=0$ automatically for the even vector field $E$ in the definition of an even or classical Jacobi structure. For odd structures this is a non-trivial condition.  Specifically, the underlying manifold $M$ is in fact a Q-manifold in the odd case.
\end{remark}
Note that the above conditions 1. and 2.  can be written entirely in terms of $S$ and $\Q$ and the canonical Poisson bracket as
\renewcommand{\labelenumi}{$\arabic{enumi}^{\prime}$.}
\begin{enumerate}
\item $\{\Q, \Q \} =0$,
\item $\{ \Q, S \}=0$,
\end{enumerate}
\renewcommand{\labelenumi}{\arabic{enumi}.}
which will be very convenient for calculational purposes.\\

In natural local coordinates $(x^{A}, p_{A})$ on $T^{*}M$ the odd Jacobi structure is given by

\begin{equation}
 S = \frac{1}{2!}S^{AB}(x) p_{B}p_{A},\hspace{15pt} \textnormal{and}  \hspace{15pt} \Q = Q^{A}(x)p_{A},
\end{equation}

 the homological vector field is in local coordinates given by $Q = Q^{A}\frac{\partial}{\partial x^{A}}$. In local coordinates the conditions on the structures can be written as

\begin{eqnarray}\nonumber
\{ \Q, \Q \} &=& 2 Q^{B}\frac{\partial Q^{A}}{\partial x^{B}} p_{A} = 0,\\
\nonumber \{\Q , S  \} &=& \left(\frac{1}{2} Q^{C}\frac{\partial S^{BA}}{\partial x^{D}} + (-1)^{\widetilde{B}} S^{BC}\frac{\partial Q^{A}}{\partial x^{c}}  \right)p_{A}p_{B}=0,\\
\nonumber \{ S,S \}+ 2 \Q S &=& (-1)^{\widetilde{C}}\left( S^{CD} \frac{\partial S^{BA}}{\partial x^{D}} + Q^{C}S^{BA}  \right)p_{A}p_{B}p_{C} = 0.
\end{eqnarray}

\begin{definition}
A manifold equipped with an odd Jacobi structure $ J:= (S,Q)$ shall be known as an \textbf{odd Jacobi manifold}.
\end{definition}

As we shall see, the algebra of smooth functions $C^{\infty}(M)$ of an odd Jacobi manifold is in fact an odd Jacobi algebra. Following the natural analogue of Lichnerowicz's constructions we have the following definition:

\begin{definition}
The \textbf{odd Jacobi bracket} on $C^{\infty}(M)$ is defined as
\begin{eqnarray}
\SN{f,g}_{J} &=&  (-1)^{\widetilde{f}+1} \{ \{ S,f \},g    \} - (-1)^{\widetilde{f}+1} \{ \Q, fg \}\\
\nonumber &=&(-1)^{(\widetilde{B}+1)\widetilde{f}  +1} S^{BA} \frac{\partial f}{\partial x^{A}} \frac{\partial g}{\partial x^{B}} + (-1)^{\widetilde{f}} \left(Q^{A} \frac{\partial f}{\partial x^{A}}   \right)g  + f \left( Q^{A}\frac{\partial g}{\partial x^{A}}  \right),
\end{eqnarray}
with $f,g \in C^{\infty}(M)$.
\end{definition}

\begin{theorem}\label{main theorem}
The odd Jacobi bracket defines an odd Jacobi algebra on $C^{\infty}(M)$. That is  the odd Jacobi bracket has the following properties:
\begin{enumerate}
\item Symmetry:  $\SN{f,g}_{J} = - (-1)^{(\widetilde{f}+1)(\widetilde{g}+1)} \SN{g,f}_{J}$.
\item Jacobi identity:  $\displaystyle\sum\limits_{\textnormal{cyclic}(f,g,h)} (-1)^{(\widetilde{f}+1)(\widetilde{h}+1)} \SN{f, \SN{g,h}_{J}}_{J}=0$.\\
\item Generalised Leibniz rule: $\SN{f, gh}_{J} = \SN{f,g}_{J}h + (-1)^{(\widetilde{f}+1) \widetilde{g}} g \SN{f,h}_{J} - \SN{f, \Id }_{J}gh$.
\end{enumerate}
\end{theorem}

\begin{proof} We proceed to prove the above theorem by making use of the local descriptions.\\
\begin{enumerate}
\item The symmetry is clear from the definition given that $S^{AB} = (-1)^{\widetilde{A}\widetilde{B}}S^{BA}$.
\item As the odd Jacobi bracket is odd and skew-antisymmetric it is sufficient to examine the \emph{even diagonal} in proving the Jacobi identity. That is we only need to consider $\SN{f,\SN{f,f}_{J}}_{J}=0$ for an arbitrary even function. Thus, via direct computation we have
    \begin{eqnarray}\nonumber
    \SN{f,\SN{f,f}_{J}}_{J} &=& (-1)^{\widetilde{C}}\left( S^{CD} \frac{\partial S^{BA}}{\partial x^{D}} + Q^{C} S^{BA}  \right)\frac{\partial f}{\partial x^{A}}\frac{\partial f}{\partial x^{B}}\frac{\partial f}{\partial x^{C}}\\
    \nonumber &-& f \left( 2 (-1)^{\widetilde{B}} S^{BC}\frac{\partial Q^{A}}{\partial x^{C}} + Q^{C} \frac{\partial S^{BA}}{\partial x^{C}}  \right)\frac{\partial f}{\partial x^{A}}\frac{\partial f}{\partial x^{B}}\\
    \nonumber &+& 2 f^{2}Q^{B}\frac{\partial Q^{A}}{\partial x^{B}}\frac{\partial f}{\partial x^{A}},
    \end{eqnarray}
collecting terms order by order in ``$\frac{\partial f}{\partial x}$".   Note that all terms involving higher order derivatives exactly cancel. Then we see that $\SN{f,\SN{f,f}_{J}}_{J}=0$ given the conditions on $S$ and $Q$ to form an odd Jacobi structure on $M$. Thus, the Jacobi identity is satisfied.
\item Via direct computation in local coordinates it is easy to see that
\begin{equation}\nonumber
\SN{f,gh}_{J} = \SN{f,g}_{J} h + (-1)^{(\widetilde{f}+1)\widetilde{g}} g \SN{f,h}_{J} - (-1)^{\widetilde{f}} Q^{A} \frac{\partial f}{\partial x^{A}}   gh.
\end{equation}
Thus the ``anomaly" is given by  $\SN{f, \Id}_{J}$.
\end{enumerate}
\proofend
\end{proof}

\begin{definition}
Given a function $f \in C^{\infty}(M)$ the associated \textbf{Hamiltonian vector field} is given by
\begin{eqnarray}
f &\rightsquigarrow & X_{f} \in \Vect(M)\\
\nonumber X_{f}(g) &=& (-1)^{\widetilde{f}} \SN{f,g}_{J} - Q(f) g.
\end{eqnarray}
\end{definition}

In natural local coordinates the Hamiltonian vector field of a function $f$ is

\begin{equation}
X_{f} = (-1)^{\widetilde{A}\widetilde{f}+1} S^{AB}\frac{\partial f}{\partial x^{B}} \frac{\partial }{\partial x^{A}} + (-1)^{\widetilde{f}}f Q^{A}\frac{\partial}{\partial x^{A}}.
\end{equation}

We will explore the properties of Hamiltonian vector fields later. Before we do this, let us examine morphisms of odd Jacobi manifolds.\\

\begin{definition}
Let $(M_{1}, S_{1}, Q_{1})$ and $(M_{2}, S_{2}, Q_{2})$ be odd Jacobi manifolds. Then a smooth map
\begin{equation}
\phi: M_{1} \rightarrow M_{2},
\end{equation}
is said to be an \textbf{odd Jacobi morphism} if and only if
\begin{equation}
\phi^{*}\SN{f,g}_{J_{2}} = \SN{\phi^{*}f , \phi^{*}g}_{J_{1}},
\end{equation}
for all $f,g \in C^{\infty}(M_{2})$.
\end{definition}

In other words, a morphism $\phi: M_{1} \rightarrow M_{2}$ is an odd Jacobi morphism if the associated pull-back morphism  is a homomorphism of odd Lie algebras.  Odd Jacobi manifolds form a category under composition of odd Jacobi morphisms.\\

\begin{proposition}
Let $(M, S, Q)$ be an odd Jacobi manifold and   $\phi : M \rightarrow M$ a diffeomorphism. Then the following are equivalent:
\begin{enumerate}
\item $\phi$ is an odd Jacobi (auto)morphism;  $\phi^{*} \SN{f,g}_{J} = \SN{\phi^{*}f, \phi^{*}g}_{J} $.
\item $\phi^{*}S = S$  and $\phi^{*}\Q = \Q$.
\item $X_{f}$ and $X_{\phi^{*}f}$ are $\phi$-related.
\end{enumerate}
\end{proposition}

\begin{proof}
$1. \Longleftrightarrow 2.$ follows from the fact that pull-back associated with the diffeomorphism $\phi$ is a symplectomorphism on the cotangent bundle and the definition of the odd  Jacobi bracket. Explicitly:
\begin{eqnarray}\nonumber
\phi^{*}\SN{f,g}_{J} &=& \phi^{*}\left( (-1)^{\widetilde{f}+1} \{ \{ S,f \},g \} - (-1)^{\widetilde{f}+1} \{\Q, fg  \} \right)\\
\nonumber &=& (-1)^{\widetilde{f}+1} \{ \{ \phi^{*}S,\phi^{*}f \},\phi^{*}g \} - (-1)^{\widetilde{f}+1} \{\phi^{*}\Q, \phi^{*}(fg)  \}.
\end{eqnarray}
Then via 1. we obtain 2.\\

$2.  \Longleftrightarrow 3.$ follows similarly. Explicitly:

\begin{eqnarray}\nonumber
\phi^{*}(X_{f}(g)) &=& \phi^{*}\left((-1)^{f} \SN{f,g}_{J} - Q(fg)   \right)\\
\nonumber &=& (-1)^{f} \SN{\phi^{*}f,\phi^{*}g}_{J} - \{ \Q, \phi^{*}(fg)  \}
\end{eqnarray}
where we have used  2. Thus,

\begin{equation}\nonumber
 \phi^{*}(X_{f}(g)) = X_{\phi^{*}f}(\phi^{*}g),
 \end{equation}
 Thus $X_{f}$ and $X_{\phi^{*}f}$ are $\phi$-related.
\proofend
\end{proof}

Taking the nomenclature from classical mechanics, we will say that an odd Jacobi automorphism is a \emph{canonical transformation} with respect to the odd Jacobi bracket.

\begin{definition}
A vector field $X \in \Vect(M)$ is said to be a \textbf{Jacobi vector field} if and only if
\begin{equation}
L_{X}S =  \{ \mathcal{X}, S \} = 0 \hspace{35pt} \textnormal{and}  \hspace{15pt} L_{X}Q = \{\mathcal{X}, \Q  \} =  0,
\end{equation}
where $\mathcal{X} \in C^{\infty}(T^{*}M)$ is the symbol or ``Hamiltonian" of the vector field $X$.
\end{definition}

The above definition is the infinitesimal version of a Jacobi automorphism. Note that the homological vector field $Q$ is a Jacobi vector field.  The Lie bracket between two Jacobi vector fields is a Jacobi vector field. The  proof follows immediately  from $L_{[X,Y]} = [L_{X}, L_{Y}]$, which itself follows from the definition $L_{X} = \{\mathcal{X}, \bullet  \}$ where $\mathcal{X} \in C^{\infty}(T^{*}M)$ is the symbol of the vector field $X$, and the Jacobi identity for the Poisson bracket.\\

\begin{lemma}\label{lemma 1}
Let $X \in \Vect(M)$ be a vector field on an odd Jacobi manifold. Then the following are equivalent:
\begin{enumerate}
\item $X$ is a Jacobi vector field.
\item $X$ is a derivation over the odd Jacobi bracket;
\begin{equation}
X(\SN{f,g}_{J}) = \SN{X(f), g}_{J} + (-1)^{\widetilde{X}(\widetilde{f}+1)} \SN{f,X(g)}_{J}.
\end{equation}
\item $[X, Y_{f}] = (-1)^{\widetilde{X}} Y_{X(f)}$, for all Hamiltonian vector fields $Y_{f}$.
\end{enumerate}
\end{lemma}

\begin{proof}
Let $X \in \Vect(M)$ be a vector field. Then consider the  symbol or ``Hamiltonian" of such a vector field: $X^{A} \frac{\partial}{\partial x^A}  \rightarrow \mathcal{X} = X^{A}p_{A} \in C^{\infty}(T^{*}M)$. The Lie derivative with respect to the vector field acting on $C^{\infty}(T^{*}M)$ is just $L_{X} = \{\mathcal{X}, \bullet \}$.\\

1. $\Longleftrightarrow$ 2.  is proved via successive use of the Jacobi identity for the canonical Poisson bracket together with 1. Explicitly:
\begin{eqnarray}\nonumber
\nonumber X(\SN{f,g}_{J}) &=&  (-1)^{(\widetilde{f} + \widetilde{X})+1} \{ \{S, \{ \mathcal{X},f \}  \},g \} - (-1)^{(\widetilde{f} + \widetilde{X})+1} \{ \Q , \{\mathcal{X},f \}g\}\\
\nonumber &+& (-1)^{\widetilde{X}(\widetilde{f}+1) + \widetilde{f}+1} \{ \{ S,f\}, \{\mathcal{X}, g  \}  \} - (-1)^{\widetilde{X}(\widetilde{f}+1) + \widetilde{f}+1} \{\Q,f \{\mathcal{X},g \} \},
\end{eqnarray}
which establishes the result.\\

1. $\Longleftrightarrow$ 3. via direct computation. Explicitly:
\begin{eqnarray}\nonumber
X(Y_{f}(g)) &=& (-1)^{\widetilde{f}} \SN{X(f),g}_{J} + (-1)^{\widetilde{f}+ \widetilde{X}(\widetilde{f}+1)} \SN{f,X(g)}_{J}\\
\nonumber &-& X(Q(f))g - (-1)^{ \widetilde{X}(\widetilde{f}+1)} Q(f)Q(g),
 \end{eqnarray}
 using the derivation property of $X$ over the odd Jacobi bracket. Then
 \begin{equation}\nonumber
 Y_{f}(X(g)) = (-1)^{\widetilde{f}}\SN{f,X(g) }_{J} - Q(f)X(g),
 \end{equation}
 gives
 \begin{equation}\nonumber
 [X,Y_{f}](g) = (-1)^{\widetilde{f}}\SN{X(f), g} - (-1)^{\widetilde{X}} Q(X(f))g,
 \end{equation}
 using 1. Thus the result is established.

\proofend
\end{proof}

\begin{corollary}\label{corollary 1}
The homological vector field $Q$ satisfies the following:
\begin{enumerate}
\item $Q(\SN{f,g}_{J}) = \SN{Q(f), g} + (-1)^{\widetilde{f}+1} \SN{f, Q(g)}$,
\item $[Q, X_{f}] = - X_{Q(f)}$,
\end{enumerate}
for all $f,g \in C^{\infty}(M)$.
\end{corollary}

\begin{proposition}\label{proposition 1}
The assignment $f \rightsquigarrow X_{f}$ is a morphism  between the odd Lie algebra on $C^{\infty}(M)$ provided by the odd Jacobi brackets and the Lie algebra of vector fields. Specifically, the following holds:
\begin{equation}
[X_{f}, X_{g}] = - X_{\SN{f,g}_{J}}
\end{equation}
for all $f,g \in C^{\infty}(M)$. Or in other words, the space of Hamiltonian vector fields is closed under the Lie bracket.
\end{proposition}

\begin{proof}
Writing out the commutator explicitly we obtain
\begin{eqnarray}\nonumber
[X_{f}, X_{g}](h)&=& (-1)^{\widetilde{f} + \widetilde{g}}\SN{f,\SN{g,h}}_{J}- (-1)^{\widetilde{f} + \widetilde{g} + (\widetilde{f}+1)(\widetilde{g}+1)}\SN{g,\SN{f,h}}_{J}\\
\nonumber &-& (-1)^{\widetilde{f}} \SN{f, Q(g)}_{J} h + (-1)^{\widetilde{g} +(\widetilde{f}+1)(\widetilde{g}+1)} \SN{g, Q(f)}_{J} h\\
\nonumber & + & \textnormal{ terms that cancel}.
\end{eqnarray}
In the above we have explicitly used the generalised Leibniz rule for the odd Jacobi bracket. Importantly all other possible terms cancel. Then using the Jacobi identity for the odd Jacobi brackets in the form
\begin{equation}\nonumber
\SN{\SN{f,g}_{J},h}_{J} = \SN{f, \SN{g,h}_{J}}_{J} - (-1)^{(\widetilde{f}+1)(\widetilde{g}+1)} \SN{g, \SN{f,h}_{J}}_{J},
\end{equation}

and the differential property of $Q$ over the odd Jacobi bracket we obtain

\begin{equation}\nonumber
[X_{f}, X_{g}] = (-1)^{\widetilde{f} + \widetilde{g}} \SN{\SN{f,g}_{J}, h}_{J} + \left(Q(\SN{f,g}_{J})\right)h,
\end{equation}
and thus the result is established.
\proofend
\end{proof}

Unlike the Schouten or indeed the Poisson case, Hamiltonian vector fields on a Jacobi manifold (both even or odd) in general do not generate infinitesimal automorphisms of the bracket structure.

\begin{proposition}\label{proposition 3}
A Hamiltonian vector field $X_{f} \in \Vect(M)$ is a Jacobi vector field if and only if $f \in C^{\infty}(M)$ is Q-closed.
\end{proposition}

\begin{proof}
If  and only if $f$ is Q-closed, i.e. $Q(f) =0$ then $X_{f} = (-1)^{\widetilde{f}}\SN{f, \bullet}_{J}$. Using Proposition(\ref{proposition 1}) we have
\begin{equation}\nonumber
[X_{f}, X_{g}] = - X_{\SN{f,g }_{J}} = (-1)^{\widetilde{f}+1} X_{X_{f}(g)},
\end{equation}
for any $g \in C^{\infty}(M)$. Then via Lemma(\ref{lemma 1}) the result is established.
\proofend
\end{proof}

Let us turn our attention to examples to show that the category of odd Jacobi manifolds is not completely empty.

\begin{example}\label{example Schouten manifolds}
\emph{Schouten manifolds}\\
These are understood as odd Jacobi manifolds for which $Q=0$. Schouten manifolds are of particular interest in mathematical physics due to their connection with the BV-antifield formalism.
\begin{enumerate}
\item Lie--Schouten structures: A vector space $\mathfrak{g}$ is a Lie algebra if and only if $\Pi \mathfrak{g}$ comes equipped with a weight minus one (``linear") homological vector field. Here $\Pi$ is the parity reversion functor.  If we employ local coordinates $(\xi^{\alpha})$ on $\Pi \mathfrak{g}$ then the homological vector field is given by:
    \begin{equation}
    Q_{\mathfrak{g}} = \frac{1}{2} \xi^{\alpha}\xi^{\beta}Q^{\gamma}_{\beta \alpha} \frac{\partial}{\partial \xi^{\gamma}}.
     \end{equation}
     A weight minus one Schouten structure on $T^{*}(\Pi \mathfrak{g}^{*})$ can be associated with the homological vector field. Employing natural local coordinates $(\eta_{\alpha}, \pi^{\alpha})$  the Schouten structure is given by
     \begin{equation}
     S = \frac{1}{2}(-1)^{\widetilde{\alpha} + \widetilde{\beta}} \pi^{\alpha} \pi^{\beta}Q^{\gamma}_{\alpha \beta} \eta_{\gamma}.
     \end{equation}
     The Schouten condition $\{ S,S \}=0$, is equivalent to the homological condition on $Q_{\mathfrak{g}}$ and in turn is equivalent to the Jacobi identity on the initial Lie algebra structure $\mathfrak{g}$.
\item Odd symplectic manifolds: An odd symplectic manifold is defined as a manifold $M$ of dimensions $(n|n)$ equipped with a closed non-degenerate odd two form denoted $\omega$, understood as a function on the total space of $\Pi TM$ . In natural local coordinates the odd symplectic form is given by
    \begin{equation}
    \omega = \frac{1}{2} dx^{A} dx^{B}\omega_{BA}(x).
    \end{equation}
    The non degeneracy condition means that, as an matrix $\omega_{BA}$ is invertible. Let us denote this inverse by $\omega^{AB}$. Then associated with an odd symplectic form is a Schouten structure given by
    \begin{equation}
    S = \frac{1}{2}(-1)^{\widetilde{B}} \omega^{AB}p_{B}p_{A} \in C^{\infty}(T^{*}M).
    \end{equation}
    The Schouten condition $\{S,S  \} =0$ is directly equivalent to the closed of $\omega$.
\end{enumerate}
\end{example}

\begin{example}\label{example q-manifolds}
\emph{Q-manifolds}\\
These are understood as odd Jacobi manifolds with $S=0$. The  odd Jacobi bracket  on a Q-manifold is given by $\SN{f,g}_{Q} = (-1)^{\widetilde{f}} Q(fg)$. Such manifold, and in particular their algebra of functions is of wide interest in mathematics due to the fact that many algebraic structures can be encoded on formal Q-manifolds.
\begin{enumerate}
\item The de Rham complex: The algebra of differential (pseudo)forms over a manifold $M$ is understood as the algebra of  functions on the manifold $\Pi TM$. The de Rham differential is the canonical homological vector field $d = dx^{A} \frac{\partial}{\partial x^{A}}$, where we have employed natural local coordinates $(x^{A}, dx^{A})$. Then over any manifold, the algebra of differential forms comes equipped with an odd Jacobi bracket given by $\SN{\alpha, \beta}_{d} = (-1)^{\widetilde{\alpha}} (d\alpha) \beta + \alpha (d\beta)$ with $\alpha, \beta \in C^{\infty}(\Pi TM)$.
\item Lie algebras: Let the vector space $\mathfrak{g}$ be a Lie algebra. Then as seen in the previous example, $(\Pi \mathfrak{g}, Q_{\mathfrak{g}})$ is a Q-manifold. Thus, $C^{\infty}(\Pi \mathfrak{g}) $ has an odd Jacobi bracket given by
    \begin{equation}
    \SN{f,g}_{Q} = \left( (-1)^{\widetilde{f}} \frac{1}{2}\xi^{\alpha} \xi^{\beta} Q_{\beta \alpha}^{\gamma} \frac{\partial f}{\partial \xi^{\gamma}}\right)g + f \left(\frac{1}{2}\xi^{\alpha} \xi^{\beta} Q_{\beta \alpha}^{\gamma} \frac{\partial g}{\partial \xi^{\gamma}}\right).
    \end{equation}
 It is worth noting that this construction of an odd Jacobi bracket generalises directly to $L_{\infty}$-algebras (c.f. \cite{Lada:1992wc}), which can be understood in terms of homological vector fields inhomogeneous in the linear coordinate $\xi$.
 \end{enumerate}
\end{example}

\begin{example}\label{example odd contact}
\emph{Odd contact manifolds}\\
Rather than diverge into the general theory of contact manifolds let us explore a specific example. Consider the manifold $M := \Pi T^{*}N\otimes \mathbb{R}^{0|1}$, where $N$ is a pure even (classical) manifold\footnote{Due to a Darboux theorem for contact manifolds this example is also generic. One can generalise the arguments of Arnold \cite{Arnold1989} (see Appendix 4) without much difficulty to include odd contact structures on supermanifolds.}. \\

Let us employ natural local coordinates $(x^{a}, x^{*}_{a}, \tau)$. The coordinates $x^{a}$ are even, while the other coordinates $x^{*}_{a}$ and $\tau$ are odd. The dimension of $M$ is $(n|n+1)$ assuming the dimension of $N$ is $n$.\\

The manifold $\Pi T^{*}N$ comes equipped with a canonical odd symplectic structure\newline  $\omega = - dx^{*}_{a} dx^{a}$.  Also note that all odd sympelctic manifolds are equivalent  to an anticotangent bundle and that the base manifold can be chosen to be a classical manifold \cite{Khudaverdian2004,Schwarz1993}.  The manifold $M$ comes equipped with an \emph{odd contact one form}, which is the even one form
\begin{equation}
\alpha = d \tau - x^{*}_{a}dx^{a}.
\end{equation}
We will for the purposes of this example stipulate that  the above is the correct form via ``superisation" of the even contact structure on $\mathbb{R}^{2n +1}$. Clearly the two form  $d \alpha = - dx^{*}_{a}dx^{a}$ is, unsurprisingly,  the canonical odd symplectic structure on $\Pi T^{*}N \subset M $.  We employ natural fibre coordinates $(dx^{a}, dx^{*}_{a}, d\tau)$ on $\Pi T M$ and  $(p_{a}, p^{a}_{*}, \pi)$ on $T^{*}M$. The coordinates $dx^{*}_{a}$, $d\tau$, $p_{a}$ are even and $dx^{a}$, $p^{a}_{*}$, $\pi$ are odd.  Then let us define the almost Schouten structure via

\begin{equation}
\phi_{S}^{*}(\alpha) =0, \hspace{15pt}  \textnormal{and} \hspace{15pt} \phi^{*}_{S}(d \alpha) = S,
\end{equation}

where we have the standard fibre-wise morphism $\phi_{S}: T^{*}M \rightarrow \Pi T M$ associated with  any almost Schouten structure.  Let us take the \emph{Ansatz}
\begin{equation}
S = p^{a}_{*}  \left( p_{a} + x^{*}_{a} \pi\right) \in C^{\infty}(T^{*}M),
\end{equation}

based on standard constructions related to even contact structures. Then

\begin{eqnarray}
\nonumber \phi^{*}_{S}(d x^{a}) &=& \frac{\partial S}{\partial p_{a}} =  p_{*}^{a}.\\
\nonumber \phi^{*}_{S}(d x^{*}_{a}) &=& - \frac{\partial S}{\partial p^{a}_{*}} = -\left(p_{a} +  x^{*}_{a} \pi  \right).\\
\nonumber \phi^{*}_{S}(d \tau) &=& -\frac{\partial S}{\partial \pi} =  -p_{*}^{a} x^{*}_{a}.
\end{eqnarray}

Direct computation gives

\begin{equation}\nonumber
\phi_{S}^{*}(\alpha) =  - p^{a}_{*}x^{*}_{a} - x^{*}_{a} p_{*}^{a} =0, \hspace{15pt}\textnormal{and} \hspace{15pt}\phi_{S}^{*}(d\alpha) = (p_{a} + x^{*}_{a}\pi)p_{*}^{a} = S.
\end{equation}

Thus our \emph{Ansatz} is confirmed to be correct. To extract the homological vector field one  needs to calculate the self Poisson bracket of the almost Schouten structure. Explicitly

\begin{eqnarray}
\nonumber \{ S,S \}_{T^{*}M} &=& 2 \frac{\partial S}{\partial p_{*}^{a}}\frac{\partial S}{\partial x^{*}_{a}}\\
\nonumber &=& 2 \left((p_{a} + x^{*}_{a} \pi)(-1)p^{a}_{*}\pi  \right)\\
 &=& - 2 (-\pi) \left(p^{a}_{*}(p_{a} + x^{*}_{a})  \right),
\end{eqnarray}

 thus  $\Q = - \pi$. It is easy to see that $\{ \Q, S \}_{T^{*}M} = 0$ and $\{ \Q, \Q \}_{T^{*}M} =0$. Furthermore, notice that the homological vector field $Q = - \frac{\partial}{\partial \tau}$ satisfies

\begin{equation}
i_{Q}\alpha = 1 \hspace{15pt} \textnormal{and} \hspace{15pt} i_{Q}(d\alpha) = 0.
\end{equation}

In short, adding an extra odd variable (``odd time") to the canonical odd symplectic manifold $\Pi T^{*}N$ produces a manifold with an odd Jacobi bracket rather than a Schouten bracket. In natural local coordinates the odd Jacobi bracket is given by

\begin{eqnarray}
\nonumber \SN{f,g}_{J} &=& (-1)^{\widetilde{f}+ 1} \frac{\partial f}{\partial x^{*}_{a}} \frac{\partial g}{\partial x^{a}} - \frac{\partial f}{\partial x^{a}} \frac{\partial g}{\partial x^{*}_{a}}\\
\nonumber &+& x^{*}_{a} \frac{\partial f}{\partial x^{*}_{a}} \frac{\partial g}{\partial \tau} - (-1)^{\widetilde{f} + 1} \frac{\partial f}{\partial \tau} x^{*}_{a}\frac{\partial g}{\partial x^{*}_{a}}\\
&+& f \frac{\partial g}{\partial \tau} - (-1)^{\widetilde{f}+1} \frac{\partial f}{\partial \tau} g.
\end{eqnarray}
\end{example}

\begin{remark}
It is worth noting that  a similar, but different generalisation of Schouten manifolds and Q-manifolds known as QS-manifolds has appeared in the literature \cite{Voronov:2001qf} (for the odd symplectic case see \cite{Alexandrov:1995kv}). The key difference is that for a QS-manifold one has a genuine Schouten structure and a homological vector field satisfying the invariance condition: $\{ S, S \} =0$ and $L_{Q}S  =0$. The odd Lie algebra structure is completely encoded in  the Schouten structure $S \in C^{\infty}(T^{*}M)$. For odd Jacobi structures the homological vector field plays an essential role in the Lie algebraic structure, one has only an almost Schouten structure $S$ satisfying a  compatibility condition with the homological vector field.
\end{remark}

\section{Maurer--Cartan elements and classical gauge systems}\label{MC elements}

In this section we mimic the BV-antifield formalism (\cite{Batalin:1981jr,Batalin:1984jr}) on an odd Jacobi manifold. We will ignore the extra gradings required  exploring only the elementary geometric and algebraic properties. Specifically, we investigate if the classical BV-antifield formalism can be modified to cope with odd brackets that do not satisfy a derivation property, i.e. not Schouten brackets.

\begin{definition}
Let $M$ be an odd Jacobi manifold. An even function \newline $s \in C^{\infty}(M)$ is said to be a \textbf{Maurer--Cartan element} if and only if
\begin{equation}
\SN{s,s}_{J}=0.
\end{equation}
\end{definition}

In more physical language, a Maurer-Cartan element is a solution to the ``classical master equation". Such an element in the BV-antifield formalism is known as an \emph{extended classical action} and governs (classical) gauge symmetries of the theory. Any solution to the ``classical master equation" is only going to be unique up to canonical transformations.    \\

\begin{definition}
Fix a Maurer--Cartan element, $s \in C^{\infty}(M) $. The \textbf{classical BRST operator} is defined as the Hamiltonian vector field associated with the Maurer--Cartan element  $s$
\begin{equation}
\delta_{s} = X_{s} \in \Vect{M}.
\end{equation}
\end{definition}

\begin{proposition}
The classical BRST operator is nilpotent. That is,
\begin{equation}
\delta_{s}^{2} = \frac{1}{2}[\delta_{s} , \delta_{s}]=0.
\end{equation}
\end{proposition}
\begin{proof}Follows directly from Proposition(\ref{proposition 1})
\begin{equation}\nonumber
[X_{s}, X_{s}] = - X_{\SN{s,s}_{J}} =0.
\end{equation}
\proofend
\end{proof}

In other words, the vector field $\delta_{s} \in \Vect(M)$ is a homological vector field. One could thus also refer to a Maurer-Cartan element  as a \emph{homological potential}, which is more adapt when discussing constructions wider than classical gauge theory.\\

In the BV formalism we require that the extended classical action \newline $s \in C^{\infty}(M)$ to
\begin{enumerate}
\item generate infinitesimal canonical transformations, i.e. the BRST operator must be a Jacobi vector field. Such transformations are identified with the BRST-gauge transformations of the theory.
\item be ``BRST-gauge invariant": $\delta_{s}s =0$.
\end{enumerate}

From the previous section it is clear that  $s$ being a Maurer-Cartan element is \emph{not} enough to ensure the above properties.  This can be remedied  by insisting that the Maurer--Cartan element be Q-closed.  \\

 Thus we are lead to the \emph{tentative} definition of a classical BV-gauge system in the context of odd Jacobi manifolds:

\begin{definition}
A \textbf{classical BV-gauge system} is the quadruple $\left(M , S, Q, s \right)$ where $\left(M, S, Q  \right)$ is an odd Jacobi manifold and $s \in C^{\infty}(M)$ is a Q-closed Maurer--Cartan element.
\end{definition}

\begin{remark}
Setting $Q=0$ reproduces the  standard definition of a classical BV-gauge system, mod the essential gradings of ghost number etc. found throughout the mathematical physics literature, see for example \cite{Lyakhovich2004} for a wide definition of a classical gauge system. For a comprehensive review of the BV formalism see the book of Henneaux \& Teitelboim \cite{Henneaux:1992ig}. Note that the properties 1. and 2. required of an extended classical action are automatically satisfied for Maurer-Cartan elements on Schouten manifolds.
\end{remark}

\section{Concluding remarks}\label{concluding remarks}
In this  paper we defined the notion of an odd Jacobi manifold, examined their basic properties and mimicked the classical aspects of the BV-antifield formalism. In particular,  it was shown that on a supermanifold equipped with an odd Jacobi structure $ J := (S, Q)$ the algebra of smooth functions over the supermanifold $C^{\infty}(M)$ comes with the structure of an odd Jacobi algebra, Theorem \ref{main theorem}. Furthermore such  the homological vector field $Q$ satisfies a derivation property over the odd Jacobi brackets Corollary \ref{corollary 1}. \\

Interestingly, following the classical aspects of the BV-antifield formalism on an odd Jacobi manifold suggest that the classical action be not just a Maurer--Cartan element but also Q-closed. Such a condition is automatic on a Schouten manifold as these can be thought of as odd Jacobi manifolds with a trivial homological vector field $Q = 0$.\\

However, it remains open as to if interesting or realistic gauge theories exist that require the use of odd Jacobi structures (with $Q\neq 0$). That said, it is conceivable that odd contact structures could find quite direct application in  theories with explicit dependency on gauge parameters.\\

The idea of ``odd time" (see Example \ref{example odd contact}.) has already been applied in the Batalin--Vilkovisky formalism to get at general and direct solutions of the master equation for a large class of gauge theories, see Dayi   \cite{Dayi1989}. In essence one understands the BRST operator as the partial derivative with respect to the ``odd time" and then  one can formulate the BV formalism in a way akin to classical mechanics. It would be very desirable to properly understand the supergeometry of Dayi's constructions and how this relates  to the work here, in  particular to odd contact manifolds.   The notion of ``odd time" is also essential when constructing flows of odd vector fields. It is certainly expected that odd contact structures are of wider interest than just their relation with odd Jacobi manifolds.  \\

The obvious areas of the present work that require further illumination include:
\begin{itemize}
\item developing the theory of odd Jacobi structures over (multi)graded manifolds and Lie algebroids. This would be required for a proper understanding of how the classical BV formalism fits into the theory of odd Jacobi structures.
\item developing the theory of higher or homotopy odd Jacobi structures and the related $L_{\infty}$-algebras.
\item understanding quantum aspects of gauge theories on  odd Jacobi manifolds via further generalisations of the BV formalism.
\end{itemize}

\section*{Acknowledgments}
The author would like to thank Prof. Janusz Grabowski  and Prof. Giuseppe Marmo for their comments and suggestions on an earlier draft of this work. 

\vfill
\begin{center}
Andrew James Bruce\\
\emph{email:} \texttt{andrewjamesbruce@googlemail.com}
\end{center}

\end{document}